\documentclass[11pt]{article}
\usepackage{graphicx}% Include figure files
\usepackage{amssymb,amsmath}
\usepackage{cite,color,url}
\def\baselinestretch{1.25}
\def\beq {\begin{equation}}
\def\eeq {\end{equation}}
\def\bea {\begin{eqnarray}}
\def\eea {\end{eqnarray}}
\def \PMET{\rm p{\!\!\!/}_T}
\def \MET{\rm E{\!\!\!/}_T}

\newcommand{\br}{\begin{eqnarray}}
\newcommand{\er}{\end{eqnarray}}
\newcommand{\be}{\begin{equation}}
\newcommand{\ee}{\end{equation}}

\begin{document}

\begin{flushright}
{IPMU13-0225}
\end{flushright}

\begin{center}
{\large \bf Study of Baryonic R-Parity Violating MSSM Using Jet Substructure Technique at the 14 TeV LHC.} 

%\vglue 0.5cm
\vskip 0.2cm
Biplob Bhattacherjee$^{a}$\footnote{biplob.bhattacherjee@ipmu.jp,
$^2$tpac@iacs.res.in},
Amit Chakraborty$^{b2}$
\vskip 0.3cm
%\\
{$^a$ 
Kavli Institute for the Physics and Mathematics of the Universe (WPI),\\
 The University of Tokyo, Kashiwa, Chiba 277-8583, Japan
}\\
\vskip 0.2cm
{$^b$  Department of Theoretical Physics, Indian Association 
for the Cultivation of Science,\\  
2A \& 2B Raja S.C. Mullick Road, Jadavpur, 
Kolkata 700 032, India}
\end{center}

%\pacs{}
\begin{abstract}
We study the discovery reach of the gluino ($\widetilde{g}$) and the lightest stop squark ($\widetilde{t}_1$) 
with baryonic R-parity violation (UDD type) in the context of Minimal Supersymmetric Standard Model (MSSM) 
at the 14 TeV run of the LHC. We consider the gluino pair production process followed by its  
decay to a top quark and a real or virtual stop squark. The top quark produced from the decay 
of the gluino can have sufficient transverse momentum to appear as a
single fat jet. We apply the jet substructure technique to tag such a hadronically
decaying boosted top quark and find that gluino mass up to 1.65 TeV can be discovered whereas
exclusion limit extends up to 1.9 TeV at the 14 TeV LHC with 300 $\rm fb^{-1}$ luminosity.
We also briefly discuss the discovery prospect of the boosted stop squark which may be identified
as a narrow resonance in the jet mass distribution.  

\end{abstract}

\section{Introduction}
The ATLAS and CMS collaborations of the Large Hadron Collider (LHC) experiment have 
announced the discovery of a Standard Model (SM) 
Higgs boson like particle with a mass of $\sim$ 125 GeV\cite{ATLAS_Higgs, CMS_Higgs}. 
They have collected 
about 25 $\rm fb^{-1}$ of data each at the end of their 7/8 TeV run, and put severe constraints 
on a large region of R-parity conserved Minimal Supersymmetric Standard Model (MSSM) parameter space. 
So far no evidence of supersymmetry (SUSY) has been found at the LHC which eventually
forces the gluino ($\widetilde{g}$) and first two generation of squarks ($\widetilde{q}$) 
to be in the TeV scale. The present bound 
on the gluino mass obtained by ATLAS collaboration in the constrained MSSM (cMSSM) is
about 1.7 TeV for $\rm m_{\widetilde{g}}\simeq m_{\widetilde{q}}$ and about 1.35 TeV for 
$\rm m_{\widetilde{g}} \ll m_{\widetilde{q}}$ where the lightest neutralino is 
assumed to be massless \cite{ATLAS_SUSY}.  
CMS collaboration also puts similar strong limits on the masses of squarks and gluino\cite{CMS_SUSY}.  
Both the ATLAS and the CMS Collaborations have searched for the  
signal of scalar partner of top and bottom quark, namely stop and sbottom, 
and placed strong limits on their masses. The CMS collaboration have analyzed 
about 19.5 $\rm fb^{-1}$ of data 
at 8 TeV center of mass energy and excluded stop mass up 
to $m_{{\widetilde t}_1} \simeq 650$ GeV \cite{CMS_stop} while 
 ATLAS collaboration ruled out stop masses about 660 GeV for a range of 
lightest neutralino masses \cite{ATLAS_stop}. 
Most of these SUSY search techniques rely on large missing transverse energy signatures originating 
from the lightest supersymmetric particle (LSP) which is stable in a R-parity conserved 
scenario and is a good candidate of non-baryonic dark matter. Applying a hard 
cut on the missing transverse energy, SM backgrounds can be significantly reduced, which 
in turn puts strong constraint on the SUSY parameter space. On the other hand, there are 
broad class of SUSY models that preserve R-parity
but lack large missing energy signatures and thus such models are less constrained compared to conventional 
SUSY models. One such possibility is the degenerate SUSY spectrum \cite{degenerate_SUSY}. As the 
exact SUSY breaking mechanism 
is still unknown to us, the possibility to have a degenerate SUSY spectrum is still an open issue. 
If such a possibility indeed exists, the current bounds on different SUSY particle masses 
at the LHC will be drastically reduced. 
For example, the gluino mass bound in such a compressed SUSY spectrum is relatively 
less constrained, about 500(600) 
GeV at the 7(8) TeV run of LHC \cite{Dreiner, BB1, ATLAS_SUSY}.  No significant improvement of 
degenerate gluino mass bound is expected at the 
14 TeV run of LHC and the limit may extend up to about 1 TeV \cite{BB2}.
Another interesting possibility is the stealth SUSY spectra \cite{Stealth_SUSY} where 
additional particles are introduced 
which leads to nearly degenerate fermion/boson pairs with small mass splitting. Current 
CMS exclusion limit on squark mass in 
stealth SUSY model is about 1.43 TeV \cite{Stealth_bound}.

As we have already discussed, supersymmetric theories with conserved R-parity provide a colorless and electrically 
neutral stable particle which  can act as a good dark matter candidate. However, the conservation of R-parity is 
not guaranteed and in case of R-parity violation, the LSP can decay to SM particles. 
R-parity is actually  connected with baryon and lepton numbers and is defined by R = (-1)$^{3B+L+2s}$, 
where B and L are baryon and lepton numbers, and s is the particle spin. All SM particle fields have R = +1 while all its
superpartner fields have R = -1. Models where R-parity is conserved, superpartners are always
produced in pairs, however inclusion of RPV interactions in the theory will allow 
single production of SUSY particles.
From the definition of R-parity, it is evident that the violation of R-parity 
would automatically mean B and/or L violation. 
However, there are certain difficulties associated with
these kind of violations unless R-parity violating interactions are sufficiently small.
Within the SM framework, both L and B are conserved quantity and so proton is 
stable with a lifetime of $\sim 10^{34}$ years. However, in the R-parity conserving 
MSSM framework the proton is stable due to absence of 
L and B violating terms in the generic MSSM superpotential. In case of R-parity violation, if both 
B and L violating terms are present, proton lifetime will become 
quite short unless those R-parity violating couplings are sufficiently small. 
Another interesting  implication of L-violating interactions arising from  
R-parity violations is the generation of neutrino masses. 
Strong constraints have already been imposed on squark/gluino masses in the context of MSSM 
with lepton number violation using multi lepton searches at the LHC \cite{RPV_lepton}. On the other hand, 
if R-parity is broken by coupling 
that violate only baryon number, bounds on squark/gluino masses are comparatively less constrained 
and we may have a chance to observe such strongly interacting MSSM particles with masses around 
1 TeV at the 14 TeV LHC.  For this reason, B violating  R-parity violation has gained recent attention both in the context 
of model building \cite{udd_model} and collider phenomenology \cite{udd_pheno}.  

In this article, we focus on the baryon number violating (UDD type R-parity violation)  scenario neglecting other forms of 
R-parity violations. The final state signature is mainly multi-jets with no (or small) missing energy which makes it the most 
challenging scenario to be searched at the LHC environment. We consider two possibilities i). gluino LSP decaying to 
top quark plus jets and ii). lighter stop squark LSP with light gluino. When the lighter stop squark is the LSP, gluino can 
directly decay to stop squark and top quark. In both cases, the final state involves top quarks with additional jets. 
Top quarks coming from the decay of gluino are generally boosted unless the mass difference between gluino and stop squark 
is small. Jet substructure technique can be very useful to identify such boosted top quark. We study the search prospect 
of gluino in the two above mentioned scenarios and also discuss the possibility to discover stop squark resonance at the 14 TeV 
LHC. The paper is arranged as follows: In Sec. 2 we briefly introduce R-parity violating MSSM and discuss current bounds on 
SUSY particles. In Sec. 3 we investigate the future prospects of gluino search performing a detailed collider simulation and 
also present our results. In Sec. 4 we study a representative benchmark point and discuss the possibility to identify a boosted 
stop squark as a resonance in the jet mass distribution. Finally in Sec. 5 we summarize our results.

\section{ R-parity violation of UDD operator and current bounds}

The superpotential of supersymmetric models with R-parity violating interactions 
($W_\mathrm{RPV}$) includes three trilinear terms parametrized by the 
yukawa couplings $\lambda_{ijk}$, $\lambda^{\prime}_{ijk}$, and
$\lambda^{\prime \prime}_{ijk}$:
\begin{equation}
\label{eqn:wrpv}
W_\mathrm{RPV} =
\frac{1}{2}\, \lambda_{ijk}L_iL_j\overline{E}_k+
\lambda^\prime_{ijk}L_iQ_j\overline{D}_k\,+\,
\frac{1}{2}\, \lambda^{\prime \prime}_{ijk}\overline{U}_i\overline{D}_j\overline{D}_k,
\end{equation}
where $i,j,$ and $k$ are generation indices; $L$ and $Q$ are the $SU(2)_L$ doublet 
superfields of the lepton and quark; and the $\overline{E}$, $\overline{D}$, and 
$\overline{U}$ are the $SU(2)_L$ singlet superfields of the charged lepton, 
down-like quark, and up-like quark.  The third term violates baryon number 
conservation, while the first and second terms violate lepton number conservation.
In this present work, we will focus on the last term and assume that lepton number is 
conserved i.e.
\begin{equation}
W_\mathrm{RPV} = \frac{1}{2}\, \lambda^{\prime \prime}_{ijk}\overline{U}_i\overline{D}_j\overline{D}_k,
\end{equation}
where $\lambda^{\prime \prime}_{ijk}$ are some arbitrary couplings that violate R-parity 
and $i,j,k$ are generation indices. Constraints on these $UDD$ type operators 
come from direct collider
searches and additionally from indirect experimental observations like $n-\bar{n}$
oscillation, renormalization group evolution etc. Detailed discussion of different indirect 
constraints on these parameters can be found in Ref.\cite{Zwirner:1984is,Goity:1994dq,
Chang:1996sw,Ghosh:1996bm}. For detailed discussions on theoretical and phenomenological 
aspects of R-parity violation 
see Ref. \cite{review}. As far as direct constraints
are concerned, the presence of R-parity violation makes the collider phenomenology 
much more complicated than the usual R-parity conserved case. Unlike R-parity conserving case,
non-zero values of $\lambda^{\prime \prime}_{ijk}$ couplings allows the LSP to decay 
to SM particles, and thus the final state signature mainly depends on the choice of 
LSP and the indices of $\lambda^{''}$ couplings. If the particle spectrum is 
such that gluino or neutralino is the LSP,
it can decay to three quarks through R-parity violating couplings.  Moderate 
constraint on gluino LSP 
comes from 3 jet resonance search by CDF collaboration of Tevatron 
experiment (144 GeV $< m_{\widetilde{g}}$) 
 ~\cite{Aaltonen:2011sg}  while  CMS collaboration of LHC experiment puts 
bound on $m_{\widetilde{g}}$  between 200 GeV and  
 460 GeV ~\cite{Chatrchyan:2012gw}. Unlike the resonance search, the 
recent study by ATLAS collaboration \cite{atlas_latest} based on counting of 
signal and background events  puts very strong constraints on gluino 
mass at the 8 TeV run of LHC with $\mathcal L = 20.3  {~\rm fb^{-1}}$  of 
 integrated luminosity assuming several possibilities as listed below: 
 \begin{itemize}
 \item $m_{\widetilde{g}} <$ 917 GeV : gluino is assumed to be the LSP and Br(gluino $\rightarrow$ light quarks(u,d,s)) = 1
 \item $m_{\widetilde{g}} <$ 929 GeV : gluino is assumed to be the LSP and Br(gluino $\rightarrow$ bottom quark ) = 1
 \item $m_{\widetilde{g}} <$ 874 GeV : gluino is assumed to be the LSP and Br(gluino $\rightarrow$ bottom, top quark ) = 1
 \item $m_{\widetilde{g}} <$ 800 GeV :  lightest neutralino is the LSP and gluino decays to neutralino assuming  
 Br(neutralino, gluino $\rightarrow$ light quarks) = 1, neutralino mass = 50 GeV
  \item $m_{\widetilde{g}} <$ 1050-1150 GeV :  lightest neutralino is the LSP and gluino decays to neutralino assuming  
 Br(neutralino, gluino $\rightarrow$ light quarks) = 1, neutralino mass = 600 GeV
   \end{itemize}

On the other hand, when lightest stop squark is the LSP, it can decay to pair 
of jets. Thus for stop pair production, the final state 
consists of 4 hard jets which can be overshadowed by large QCD background. 
So far no bound on stop squark mass when it decays 
to jets exists \cite{Evans:2012bf}.  However, if the final state contains b jets, 
the SM background may be small and there is a chance to discover 
stop squark with mass up to  200 GeV with 20 $\rm fb^{-1}$ of integrated 
luminosity at the 8 TeV LHC \cite{Franceschini:2012za}. 
At the 14 TeV run of LHC, the expected discovery reach of stop squark 
with multi-jet final state is about 800 GeV  with 300 $\rm fb^{-1}$ 
data \cite{Duggan:2013yna}.

We can see from the above discussion that the current limit on gluino mass is 
less than 1 TeV in UDD type R-parity violating 
scenario and the bounds depend on specific choices of gluino decay modes.  
Additionally stop squark mass $\sim$ 100 GeV 
is still allowed if it decays to a pair of jets. This implies that there are 
several possibilities where the bounds are not directly applicable, providing 
the opportunity to perform further studies in this direction.  

\section{Details of collider simulation and results}

We already discussed that in the presence of R-parity violating (RPV) couplings 
we have the freedom to choose the 
lightest supersymmetric  particle (LSP) and its decay to various SM particles.  
Here we consider a simplified version 
of RPV MSSM  
where the lighter stop squark ($\widetilde{t}_1$) and the gluino ($\widetilde{g}$) 
are within the reach of LHC while other SUSY 
particles are relatively 
heavy. Thus, we have the option to 
choose either the lighter stop squark or gluino as the LSP candidate. 
When stop is the LSP candidate, then gluino will decay to a top quark and stop 
squark ($\widetilde{t}_1$). Note that, 
in order to allow the decay of $\widetilde{t}_1$ to SM particles, the 
first index of $\lambda^{''}_{ijk}$ coupling should be equals to 3
(i.e., $\lambda^{''}_{3 jk}$) and the corresponding decay of $\widetilde{t}_1$ is
\begin{eqnarray}
\widetilde{t}_1 \rightarrow  j j  
\label{e1}
\end{eqnarray} 
 where one of the jets may be a  b-jet depending upon the choice of indices j and k. 
This leads to the final state from the decay of gluino as follows
\begin{eqnarray}
\tilde{g} \rightarrow t \widetilde{t}_1 \rightarrow t j j. 
\label{e2}
\end{eqnarray} 
Here we assume that the mass difference between the gluino and stop squark is always greater 
than top quark mass. 

On the other hand, when gluino is the LSP and the assumed RPV coupling is $\lambda^{''}_{3 jk}$, 
then gluino will decay via three body final states i.e.,
\begin{eqnarray}
\tilde{g} \rightarrow t j j  
\label{e3}
\end{eqnarray} 
On a passing note, since gluino is a majorana particle, we may get same 
sign top quarks
from the gluino pair production and thus final state may contain same 
sign di-leptons. The expected sensitivity of 14 TeV LHC with high luminosity 
option in this channel 
have already been discussed~\cite{Saelim:2013gea}. However, if the top quarks coming 
from the decay of gluino is highly boosted, 
the decay products of top quark are not generally isolated. As a result, the 
probability to have final states involving 
isolated leptons are relatively small compared to the conventional scenario 
with low $p_T$ top quarks. 
Here we focus on the gluino pair production
process with gluino decaying via both the two body (Eqn.~\ref{e2})
and three body (Eqn.~\ref{e3}) decay modes, provided 
they are kinematically allowed.
\begin{eqnarray}
p p \to \tilde{g} \tilde{g} \to t t + jets 
\label{e4}
\end{eqnarray} 
Hence, the final state signature includes at least two top quarks 
with additional jets. Throughout our analysis we assume that the $\lambda^{''}$ couplings lead to short enough lifetimes of 
gluino and stop squark such that they decay promptly. For a gluino with mass in the TeV regime, the top quark
produced from the gluino decay can have sufficient transverse momentum
to appear as a single fat jet. One can thus apply the Jet substructure 
technique to reconstruct the invariant mass of the top quark. Similarly, the stops
may also have sufficient transverse boost to appear as a fat jet. In the next 
section, we explore the possibility to have a boosted stop squark and 
find that the jet substructure is a useful technique to reconstruct
the stop from the busy LHC environment. There are SM processes
which can give rise the final state signature we are interested in and thus 
contribute to the background for our signal. In our analysis, we consider 
the two most important backgrounds namely $\rm t \bar{t} +  jets$ (up to 2 jets)
and QCD jets which can play a significant role in our region of interest\footnote{We also consider 
single top quark production ( top + b quark , top + light quark and top + W)  and we find that it can be reduced to a negligible level after imposing our final optimized
selection cuts. }. As our final 
state signature do not include any lepton, we require special attention
to reduce the contamination of the QCD background thereby enhancing the 
signal significances. We use PYTHIA6.4.24\cite{Sjostrand:2006za} for
the generation of the signal events while MadGraph5 \cite{Alwall:2011uj} generates
the background events and subsequently the MadGraph-PYTHIA6 interface is used 
to perform the parton showering and implement
our event selection cuts. We include the matching of the matrix element hard partons and
shower generated jets following the MLM prescription\cite{Hoche:2006ph}. 
We use the FASTJET (version 3.0.4)
\cite{Cacciari:2011ma} for the reconstruction
of jets and the implementation of the jet substructure analysis for
the reconstruction of the top quark. 
As we already mentioned, it is always a challenging task to get rid of the QCD backgrounds when the signal 
do not contain any lepton. On the other hand,  we expect hard jets coming from the decay of massive gluino (TeV scale)  
or stop squark and the scalar sum of the  transverse momentum of final state visible particles should be 
 above 1 TeV. Keeping this in mind, we proceed to generate the event sample for QCD multi-jets for four different region of
the effective mass ($S_T$) variable namely $S_T >$ 1200~GeV, 1200~GeV $< S_T < 2000$~GeV, 
$S_T >$ 2000~GeV, and finally $S_T >$ 3000~GeV where $S_T$ is defined as follows:
\begin{equation}
S_T = \displaystyle\sum_{j=1}^4 p_T^j. \nonumber
\end{equation}
%In order to focus on the high $S_T$ regime,
%we set the minimum threshold for the QCD jets at 200 GeV.
For wide separation of gluino and stop squark masses, both stop squark and top quark may appear as fat jets and 
we can expect four fat jets at the parton level. That is why we are interested in four high $p_T$ fat jets in the final state. 
While simulating the QCD background events, we restrict ourselves
up to four jets at the parton level due to our computational limitations. However, the 
QCD event sample includes events with higher ($>$ 4) jet multiplicities 
as we allow parton showering of 4-jet samples.
For the $t\bar{t}$+ jets, we also generate sufficient number of events for both low and high $p_T$ samples to  
cover the entire phase space.

  Though the final state signature we are interested in includes at least one
top quark, however to make our analysis more general, we consider three possibilities 
a) final state with no tagged top quark jet, b) at least one tagged top quark jet and 
c) at least two tagged top quark jet. We tag the top quarks using the publicly 
available Heidelberg-Eugene-Paris Top-tagger (HEPTopTagger) \cite{Plehn:2010st,Plehn:2011tg} 
package with its default settings assuming that the top 
quark decays hadronically to produce jets in the final state.
The top-tagging technique is primarily based on the Cambridge-Aachen (C/A) \cite{Dokshitzer:1997in} jet algorithm
and a mass drop criteria along with a filtering technique (for details, see\cite{Plehn:2011tg,Butterworth:2008iy}. 
We also use C/A algorithm with jet radius 
R = 1.2 to reconstruct the fat jets and then fed these jets as the input 
of the top-tagging algorithm.

We will now describe the details
of our simulation procedure as well as the kinematic cut optimization 
technique opted to enhance the signal significances. Before we proceed, we would like to
discuss the choice of the parameters that are involved in our analysis. 
The values of $\rm M_{1}$ and $\rm M_{2}$ are set to 5 TeV
as these are irrelevant for the parameter space of our interest. The higgsino mass 
parameter $\mu$ is taken to be 5 TeV and tan$\beta$, the ratio of the vacuum expectation 
values of the two Higgs doublets, is fixed at the value of 5. 
The masses of the first two generation of 
squarks and all the three generation of sleptons along with the right handed third 
generation squark (sbottom) mass parameter are also set to 5 TeV as 
they will not play any significant role in our study. All the tri-linear 
couplings $\rm A_t$, $\rm A_b$ and $\rm A_{\tau}$ are set to zero as these  
couplings have very little impact on our analysis. As we discuss, we are 
mainly interested in the decay of gluino to a top quark, the two important 
parameters that have significant impact on these decay modes are gluino mass parameter 
$\rm M_{3}$ and the right handed third generation squark (stop) mass parameters 
$\rm m_{\widetilde{t}_{R}}$. We perform a dedicated scan in the $M_{\widetilde{g}}$ -- $ M_{\widetilde{t}_{1}}$ 
plane in order to understand the sensitivity of our search strategy over a wide range. 
So, we start with the gluino mass from 500 GeV to 2 TeV,
while stop mass varies from 100 GeV to 2 TeV with smaller bin sizes ($\sim$25/50 GeV). 
we use Prospino2.1\cite{prospino} to calculate the gluino pair production cross section at the 
next-to-leading order at the 14 TeV LHC. Additionally, we also study the three body decay mode 
of gluino i.e. $\tilde{g} \to t j j$, varying the gluino mass from 500 GeV to 2 TeV in step 
of 50 GeV fixing $\rm M_{\widetilde{t}_{1}}$ at 3 TeV while other parameters 
are same as mentioned above. We use the top quark mass 173.1 GeV in our whole
analysis \cite{top_mass}.

Our final state topology do not include any isolated lepton, hence we will not 
consider any observable associated with the kinematics
of the lepton. Besides, we consider the hadronically decaying top quark,
thus there is no lepton and no real source of missing energy ($\MET$) like 
the neutrinos. Hence, we will not consider $\PMET$ as a relevant kinematic 
observable in our analysis, rather the most interesting observables are actually 
the transverse momentum of the jets and the effective mass ($S_T$) constructed 
using the transverse momentum of the jets.

The basic idea behind this cut optimization technique is to find out 
the possible combinations of the five relevant observables 
namely $p_T^{j_1}$, $~p_T^{j_2}$, $~p_T^{j_3}$,$~p_T^{j_4}$ and $~S_T$
such that signal significance ($\mathcal S$) takes the maximum value.
We vary these observables in the following ranges with small step 
sizes ($\sim$ 50/100 GeV).    
\begin{eqnarray}
100< p_T^{j_1} < 1000, &\ \ 100< p_T^{j_2} < 1000, &\ \ 50< p_T^{j_3} < 500, \nonumber \\
50< p_T^{j_4} < 500, &\ \ 1200< S_T < 3500
\label{param1}
\end{eqnarray}  

For each point corresponding to this five dimensional parameter space,
we calculate the signal significance ($\mathcal S$) as defined below using 
the signal and combined background events, 
\begin{equation}
\mathcal S = \frac{\rm N_S}{\sqrt{\rm N_B + (\kappa N_B)^2}} \, \, ,
\end{equation}
where $\rm N_S$  and $\rm N_B$ are the number of signal and background events respectively and 
$\kappa$ is the measure of the systematic uncertainty. We assume $\kappa$ = 20\% and 
$\mathcal L $ = 300 $\rm fb^{-1}$ of luminosity at the 14 TeV run of LHC.
We calculate $\mathcal S$ for nearly $\sim$ 53000 possible combination of these observables 
for zero, one and two  top-tagged samples separately.
We choose a few optimized set of kinematic cuts to discuss the signal and background
characteristics, as shown in Table \ref{tab1}. We define the discovery and exclusion limits if $\mathcal S$ is greater than 
5 and 2 respectively for a particular choice of gluino/stop mass.  
We denote the signal regions as SRXn where X is an arbitrary label corresponding to various set of 
cuts (X $\equiv$ A-C)  and the quantity n here denotes the number of
tagged top quark. 

We find that the signal involves sizeable number of events with jet multiplicity greater than four. 
For this reason we further consider events with five and six jets and take approximately 50 possible combinations 
of the jet transverse momentum and $S_T$ around the four jet optimized cuts. We would like to remind our readers 
that our 5-jet and/or 6-jet limits may not be so robust compared to 4-jet limits due to our inability to generate 
5-jet and 6-jet background samples at the parton level, still they can be considered as an approximate estimation of the 
discovery reach at the LHC. 

We get two distinct signal regions for the one 
top-tagged sample. For large values of gluino masses SRA1 helps 
to get the 5$\sigma$ discovery reach for 1-toptag sample which includes five jets while 
for relatively small values of gluino masses a cut set similar to SRB2 helps. 
For the two top-tagged sample, the signal region SRC2 is the most efficient cut set with highest LHC discovery
reach. On the other hand, the discovery limit for the zero top-tagged case is much weaker than that of 
one and/or two top-tagged scenario. This means top-tagging is quite efficient for the reduction of 
backgrounds while keeping sufficient number of signal events thereby enhancing the signal significance. 

\begin{table}[ht]
\centering
\begin{tabular}{|c|ccccccc|c|}
\hline
& $P_T^{j_1}$ & $P_T^{j_2}$  & $P_T^{j_3}$ & $P_T^{j_4}$ & $P_T^{j_5}$ & $P_T^{j_6}$
& $S_T$ & \# of top-tag \\
\hline\hline
SRA1 & 400 & 400 & 400 & 400 & 400 & -- & -- & 1\\ [-2mm]
SRB2 & 300 & 200 & 200 & 200 & -- & -- & -- & 2\\ [-2mm]
SRC2 & 800 & 600 & 500 & 400 & -- & -- & 2500 & 2\\ 
\hline
\end{tabular}
\def\baselinestretch{1.1}
\caption{\small {\it Details of three optimized kinematic cuts
which provide the maximum sensitivity in the search
for stop ($\widetilde{t}_1$) and gluino ($\widetilde{g}$). The last column indicates the number of tagged top quarks
in the final state. All the jet momenta along with the effective mass parameter ($S_T$) are expressed in units of GeV.} }
\label{tab1}
\end{table}

In Table ~\ref{tab2}, we display the number of events
survive after each cut set for both the signal and the 
background events. To describe the signal event, we
consider a representative benchmark point (BP-1) where
the gluino mass is fixed at 1.2 TeV and the 
lighter stop mass at 550 GeV keeping other parameters
same as before.

%............................................................
%
\begin{table}[htbp]
\centering
\begin{tabular}{|c|ccc|}
\hline
 & SRA1 & SRB2  & SRC2 \\
\hline\hline
QCD jets & 12  & 30  & 30  \\ %[-2mm]
\hline
$t \bar{t}$+jets & 22  & 72  & 15  \\ 
\hline
Total bkg & 34 & 102 & 45  \\
\hline\hline
BP-1 & 49 & 180 & 58  \\ %[-2mm]
\hline\hline
Significance ($\mathcal S$) & 5.5 & 7.9 & 5.2 \\ %\\[-2mm]
\hline
\end{tabular}
\def\baselinestretch{1.1}
\caption{\small {\it Number of events
survive after each cut set (see Table \ref{tab1})
for both the signal and the background events.
We consider a sample benchmark point (BP-1) with
$M_{\widetilde{g}}$ = 1.2 TeV and $M_{\widetilde{t}_1}$ = 550 GeV.
We assume $\mathcal L $ = 300 $\rm fb^{-1}$ with
20\% systematic uncertainty at 14 TeV run of LHC.}}
\label{tab2}
\end{table}

%............................

\begin{figure}[t!]
\begin{center}
\includegraphics[scale=0.4]{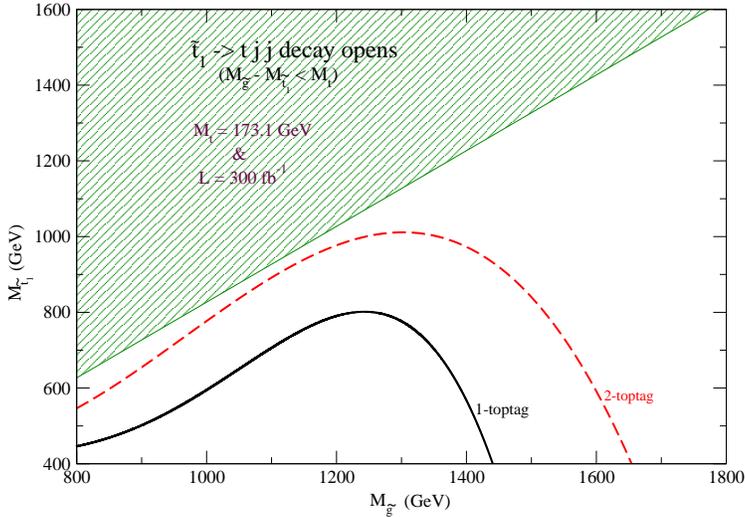}
\caption[]{\small{\it The 5$\sigma$ discovery limit in the 
($M_{\widetilde g}-M_{\widetilde{t}_1}$)
plane for UDD type R-parity violating MSSM. The diagonal 
line $M_{\widetilde{g}} - M_{\widetilde{t}_1} = M_t$ separates
two region with distinct kinematics. The solid black line represents
the one top-tagged limit, while dashed red (light grey) denotes the 
limits for 2 top-tagged sample.
We assume $\mathcal L = 300~ {\rm fb^{-1}}$ luminosity
and 20\% systematic uncertainty at the 14 TeV LHC.}}  
\label{fig:mst_mgl}
\end{center}
\end{figure}

%............................
In Fig \ref{fig:mst_mgl}, we display the projected discovery reach in 
the $M_{\widetilde{t}_1} - M_{\widetilde{g}}$ plane in the context of 14 TeV run of 
LHC. The straight (blue)  line $M_{\widetilde{g}} - M_{\widetilde{t}_1} = M_t$ separates
two regions with distinct kinematics, below which the gluino decays 
in two body final state while in the above region three body decay mode is important.
Near this line, the stop ($\widetilde{t}_1$) and top are produced almost at rest
and thus the transverse momentum of top quark is very small, thereby
reducing our acceptance limit. We find that SRC2 gives the best sensitivity
where we tag two top quarks, however the limit obtained from SRA1 
(signal with one top tag) is slightly weaker. We also find that the gluino mass as high 
as 1.4 TeV is achievable for stop mass $\sim$ 1 TeV with 300 $\rm fb^{-1}$ of data at 
the 14 TeV LHC. It is to be noted that, in the case of zero top-tagged sample, 
one can achieve 5$\sigma$ signal significance for gluino masses upto 800 GeV
at the 14 run of LHC. For the region above the straight (blue) line, the gluino dominantly decays
to top quark and light jets. The 5$\sigma$ reach in the gluino mass is approximately 1.4 TeV 
for 300 $\rm fb^{-1}$ of luminosity, the limit do not dependent on the choice of stop mass. 
We also study the sensitivity of our results considering 50 \% enhancement in total SM background 
estimation and find that the discovery reach is reduced by about 150 - 200 GeV. The gluino exclusion 
limits for the above mentioned two possibilities are about 1.9 TeV (gluino 2 body decay) and 1.7 
TeV (gluino 3 body different) respectively. 

Before we conclude this section we would like to comment on some important issues: 
\begin{itemize}
\item In addition to light jets, b jets may also appear in the final state either 
from the decay of top or from stop decay. In our analysis we do not identify b-jets 
in the final state rather we assume these as inclusive jets.  

\item  Improvement can be expected with the High Luminosity LHC (HL-LHC) run, with 3000 $\rm fb^{-1}$ of 
luminosity if we impose harder cuts on $p_T$ and $S_T$ of the jets. However, with such high luminosity, 
pile up is an important issue which can increase the systematic uncertainty. In that case the discovery limits 
on gluino and stop masses may not change significantly.  

\item Although we assume that the strength of $\lambda^{''}$ coupling is such that gluino and stop decay promptly, 
the lifetime of these particles can be large enough. This may lead to displaced vertices, stable gluino/stop for collider 
etc. In order to study such signals, dedicated analysis taking care of detector effects are required which is beyond the 
scope of this paper.

\item Triggering is an important issue in R-parity violating searches at the LHC. 
CMS collaboration have studied the signatures of multijets in the context of R-parity 
violating MSSM using both 7 TeV and 8 TeV data \cite{Chatrchyan:2012gw,Chatrchyan:2013gia}, where the
trigger used to select signal events are based on the scalar sum of the transverse 
momentum of all the jets ($H_T$) measured using calorimeter information. 
In addition, to reduce the effects of multiple pp interactions, the jets 
are selected with $p_T >$ 40 - 60 GeV. Here, we study the discovery reach at the LHC in the presence of a 
number of energetic jets in the final state in the context of RPVMSSM.
A trigger, very much similar to that of the CMS collaboration, as discussed above, 
will be sufficient enough to select our signal events. We check that 
a choice of $H_T >$ 1000 GeV as our trigger with jets having $p_T >$ 50 GeV, will select 
almost 99\% of our signal events.

\end{itemize}

%%%%%%%%%%%%%%%%%%%%%%%%%%%%%%%%%%%%%%%%%%%%%%%%%%%%%%%%%%%%%%%%%%%%%%%%%%%%%%%%%%%%%%%%%%%%%%%%%%%

\section{ A simplified benchmark study}

In this section, we discuss the possibility to discover the lighter 
stop squark at the LHC performing a simplified benchmark study.
Gluinos are pair produced at the LHC followed by a direct decay of gluino 
to a top quark and a stop squark and the stop squark decays 
to a pair of jets. We assume a wide separation of mass between
the stop and gluino such that the stop squarks produced from the decay
of gluino will be sufficiently boosted. The most important feature of 
a boosted particle decaying into multiple hadronic jets is that the 
final states remain highly collimated because of large transverse
momenta of the parent particle, thus appearing as a single fat jet. 
Conventional jet finding algorithms will have less sensitivity in such cases, this 
realization has led to the idea of {\it{Jet Substructure}}\cite{Butterworth:2008iy} 
technique.\footnote{For direct stop pair production, one can also apply jet substructure 
technique. For details, see Ref.~\cite{stop_jet}.} Interestingly, the situation is very much similar to the 
Higgs decay to a pair of b-jets which also appear as a single fat jet when 
the Higgs has relatively large transverse momentum. The BDRS Higgs 
tagger \cite{Butterworth:2008iy} which is based on the jet substructure 
technique was developed primarily to study the boosted Higgs fat jets.
Here we apply this Higgs tagging technique to look for the 
signatures of the stop resonances at the LHC. Our stop
tagging algorithm is as follows:

\begin{itemize} 
\item  Final state jets (from $ {\widetilde{t}_1} \to j j $ ) are formed based 
on an Cambridge Aachen jet clustering algorithm with jet
radius R = 1.2.   
\item We investigate the sub-jet kinematics step by step and apply the mass drop criteria
along with the filtering technique at the sub-jet level. 
\item Select the best sub-jets to form the fat jet mass,  
which essentially should correspond to the parent particle mass.
\end{itemize}

\begin{figure}[ht!]
\begin{center}
\includegraphics[scale=0.4]{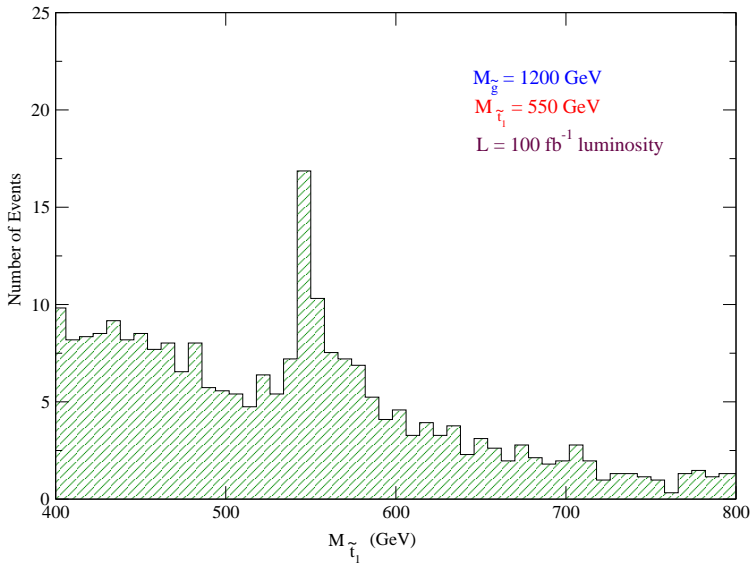}
\caption[]{ \small {\it Invariant mass distribution of the
reconstructed jets using the boosted jet substructure 
technique for a sample benchmark point. 
Here we assume $\rm M_{\widetilde{t}_{1}}$ = 550 GeV
and $\rm M_{\widetilde{g}}$ = 1.2 TeV. We see clear resonance peak
with 100 ${\rm fb^{-1}}$ luminosity at the 14 TeV run of LHC.}}
\label{fig:mstreco}
\end{center}
\end{figure}

We fix the gluino mass parameter $\rm M_{\widetilde{g}}$ at 1.2 TeV and the 
stop mass $\rm M_{\widetilde{t}_1}$
at 550 GeV keeping other parameters fixed at the values as mentioned in Sec. 3. 
In Fig~\ref{fig:mstreco}, we show the jet mass distribution reconstructed using the above 
mentioned substructure algorithm. From the figure
it is clear that we get a resonance peak around 550 GeV which is 
indeed equal to the mass of the lightest stop (${\widetilde{t}_1}$)
present in this event. Here we assume $\mathcal L $ = 100 $\rm fb^{-1}$ of 
luminosity at the 14 TeV run of the LHC. One should note that the width of stop squark is small compared to 
detector resolution as the RPV coupling is very small. However, we get sizeable width in the 
jet mass distribution due to the presence of various effects like smearing, multiple interaction etc. 
One may expect to observe a resonance peak 
at/around the gluino mass using the invariant mass distribution of the 
reconstructed top and stop. However, it is not possible to have such 
a resonance like structure as gluino decay width is mostly
determined by the QCD couplings and it is always very large, resulting 
into a broad jet mass distribution. One should also note that the resonance peak 
will disappear for small mass difference between gluino and stop squark due to small boost of the 
produced stop squarks.

\section{Conclusions}
Supersymmetric theories with conserved R-parity naturally
provide a stable lightest SUSY particle (LSP) which
can be a good dark matter candidate. However, in the
presence of R-parity violating couplings in the
superpotential, the LSP can now decay to SM particles
which leads to final state signatures with small
missing transverse energy. For example, the UDD type
of R-parity violation allows the LSP to decay to
 multiple jets which makes it a very challenging scenario
to search at the LHC environment. In this paper, we focus on such
UDD type of R-parity violation assuming two cases a) stop ($\widetilde{t}_1$) is the LSP
or b) gluino ($\widetilde{g}$) is the LSP. We further
assume the multi-jet final states originating
from the decay of the gluino contains top
quark with sufficient transverse momentum. We apply
the jet substructure technique to reconstruct these
hadronically decaying top quarks and calculate discovery as well 
as exclusion limits on the mass of the gluino. We find
that gluino masses up to 1.65 TeV can be discovered with
300 $\rm fb^{-1}$ of luminosity at the 14 TeV LHC run.
With higher luminosities, one may expect some improvement in 
the discovery reach on the gluino mass, however it is to be noted 
that pile up will play significant role with such high luminosities.
We also discuss how jet substructure technique can be very useful to search 
for the resonance peak of the stop squark ($\widetilde{t}_1$) in the early run of
14 TeV LHC. This method to identify stop squark as a single fat jet is efficient 
only if  the mass gap between gluino and stop squark is large.

\section{Acknowledgements}
A.C would like to thank Kavli Institute for the Physics and Mathematics 
of the Universe (Kavli IPMU), Japan where initial part of the work was 
done. A.C also thanks Sutirtha Mukherjee and Ipsita Saha for discussions 
and help in preparing some of these figures using Xmgrace. 
B.B acknowledges the support of the World Premier International 
Research Center Initiative (WPI Initiative), MEXT, Japan and
would also like to thank Indian Association for the Cultivation of Science (IACS), 
India for the hospitality during the final stage of this work.
A.C sincerely thank Dilip Kumar Ghosh for his constant encouragement and support.

%%%%%%%%%%%%%%%%%%%%%%
%%%%% References %%%%%
%%%%%%%%%%%%%%%%%%%%%%

\end{document}